\documentclass{ifacconf}

\usepackage{graphicx}      
\usepackage{wrapfig}
\usepackage{natbib}        
\usepackage{amsmath} 
\usepackage{amssymb}  
\usepackage{amsfonts}
\usepackage{tikz}
\usepackage{caption}
\usepackage{subcaption}

\newcommand{\dfb}{\stackrel{\Delta}{=}}
\newcommand{\R}{\ensuremath{\mathbb R}}

\newtheorem{remark}{Remark}

\begin{document}
\begin{frontmatter}

\title{Distributed algorithm for controlling scale-free polygonal formations.\thanksref{footnoteinfo}} 

	\thanks[footnoteinfo]{The work of Hector Garcia de Marina was supported by Mistrale project, http://mistrale.eu. The work of Cao was supported in part by the European Research Council (ERC-StG-307207) and the Netherlands Organization for Scientific Research (NWO-vidi-14134).}

\author[First]{Hector Garcia de Marina}
\author[Second]{Bayu Jayawardhana}
\author[Second]{Ming Cao}

\address[First]{University of Toulouse, Ecole Nationale de l'Aviation Civile (ENAC) Toulouse 31000, France (e-mail: hgdemarina@ieee.org).}
	\address[Second]{Engineering and technology institute Groningen (ENTEG), University of Groningen, the Netherlands (e-mail: \{b.jayawardhana,m.cao\}@rug.nl}

\begin{abstract}                
This paper presents a distributed algorithm for controlling the deployment of a team of mobile agents in formations whose shapes can be characterized by a broad class of polygons, including regular ones, where each agent occupies a corner of the polygon. The algorithm shares the appealing properties of the popular distance-based rigid formation control, but with the additional advantage of requiring the control of fewer pairs of neighboring agents. Furthermore, the scale of the polygon can be controlled by only one pair of neighboring agents. We also exploit the exponential stability of the controlled formation in order to steer the formation as a whole with translations and rotations in a prescribed way. We provide both theoretical analysis and illustrative simulations.
\end{abstract}


\end{frontmatter}

\section{Introduction}
Accomplishing the tasks of surveillance, exploration, and search \& rescue, among others, often requires \emph{formation control} of multi-agent systems (see for instance \cite{oh2015survey}). In particular, an appealing formation control framework based on \emph{rigid frameworks} for the above mentioned tasks has been discussed in \cite{KrBrFr08}. In such setups the agents can form a desired shape by only controlling the distances between neighbors. It is worth mentioning as follows some of the properties of distance-based rigid formation control. Firstly, the agents do not need to share a common frame of coordinates. Secondly, the system is robust against biases when sensing the distances between of neighboring agents (see \cite{MarCaoJa15}). Thirdly, the motion of the formation can be de-coupled into rotational, translational and \emph{scaling} movements (we refer to \cite{Hector2016maneuvering,HectorCDC16}). Forthly, the stability of the closed-loop system is \emph{exponentially stable} for agents modelled by first or second-order integrators \cite{sun2016exponential}. On the other hand, the main drawback of this distance-based approach is that to control the formation, one needs to control at least $2n - 3$ distances in 2D, in order to be able to achieve a desired shape. This is not required for other approaches such as the position-based control \cite{oh2015survey}, but with the possible price of loosing many of the above listed advantages.

This paper presents an algorithm for controlling formations in the shapes from a broad class of polygons, i.e., a plane figure that is bounded by a finite chain of straight line segments closing in a loop to form a closed chain, where each agent occupies a corner of the polygon. We will show that the algorithm has all the advantages from distance-based rigid formation control and at the same time it only needs to control a smaller number of agent pairs. In particular, the assignment of neighboring agents, e.g., the sensing topology of the team, is based on a daisy chain configuration, i.e., a configuration where the agents are connected in series. We will also show that by controlling only the distance between the first and last agent in the sensing topology, one can scale the size of the whole formation up and down freely.

The algorithm is motivated by the distance-based control of non-rigid formations as recently studied in \cite{dimarogonas2008stability}. In particular, we exploit the effect of having mismatches in the prescribed distances between neighboring agents. Although one cannot define a particular shape by controlling a non-rigid setup, it is reported in \cite{ifacnew} that biases in the range sensors of neighboring agents\footnote{In the cited paper, the mismatches have been addressed as a biases. Nevertheless, mathematically speaking in the cited paper both concepts are equivalent.} can be used to lead the formation to converge to a collinear configuration for a daisy chain network consisting of three agents. In this work we will employ a slightly different approach than in \cite{ifacnew}. In fact, we will strengthen the resultant \emph{mismatched} control law by clearly identifying two terms. The first term is responsible for controlling distances and derived from the standard gradient descent technique for the chosen potential function. The second term involving the mismatches has a clear interpretation and is responsible for the steady-state collinear configuration. Furthermore, as hindsight, the first term is surprisingly in the same form as the control law presented in \cite{kvinto2012equidistant} and \cite{proskurnikov2016problem} for steering equally-spaced agents to a line. We will show that with the technique introduced in \cite{ifacnew}, the second term responsible for the alignment of the formation can be modified in order to control a prescribed angle and a prescribed distance between two pairs of consecutive neighboring agents in the daisy chain. Furthermore, the scale of the whole formation can be set by one pair of neighboring agents. The proposed algorithm renders the controlled formation convergning to the prescribed shape \emph{exponentially stable}. This property, combined with a non-fixed steady-state orientation, allows us to achieve translations and rotations of the desired shape by following the technique introduced in \cite{Hector2016maneuvering}.

This paper is organized as follows. We introduce some notation and the notion of framework in Section \ref{sec: not}. Then in Section \ref{sec: dai} we introduce the daisy chain topology for distance-based control. The addition of distance mismatches between neighboring agents in the control terms leads to an algorithm for deploying agents in a collinear and equally (or relatively) spaced configuration. We modify this algorithm by the addition of rotational matrices in Section \ref{sec: rot} in order to control the relative angle between two consecutive relative positions in the framework. We prove the exponential stability under the new algorithm for a broad class of polygons, including regular ones. At the end of Section \ref{sec: rot} we exploit this exponential stability in order to control the scale of the desired shape by only controlling the distance between the first and the last agent of the framework, and to induce rigid body motions, i.e., rotations and translations, to the polygon. We present a numerical simulation in Section \ref{sec: sim} in order to validate the theoretical findings and we finish the paper with some conclusions in Section \ref{sec: con}.

\section{Notations and definitions}
\label{sec: not}
For a given matrix $A\in\R^{n\times p}$, define $\overline A \dfb A \otimes I_2 \in\R^{2n\times 2p}$, where the symbol $\otimes$ denotes the Kronecker product and $I_2$ is the $2\times 2$ identity matrix. We denote by $|\mathcal{X}|$ the cardinality of the set $\mathcal{X}$.

Consider a formation of $n\geq 3$ autonomous agents whose positions are denoted by $p_i\in\R^2$ with $i\in\{1,\dots,n\}$. The agents are able to sense the relative positions of their neighboring agents. The neighbor relationships are described by an undirected graph $\mathbb{G} = (\mathcal{V}, \mathcal{E})$ with the vertex set $\mathcal{V} = \{1, \dots, n\}$ and the ordered edge set $\mathcal{E}\subseteq\mathcal{V}\times\mathcal{V}$. The set $\mathcal{N}_i$ of the neighbors of agent $i$ is defined by $\mathcal{N}_i\dfb\{j\in\mathcal{V}:(i,j)\in\mathcal{E}\}$. We define the elements of the incidence matrix $B\in\R^{|\mathcal{V}|\times|\mathcal{E}|}$ for  $\mathbb{G}$ by
\begin{equation}
	b_{ik} \dfb \begin{cases}+1 \quad \text{if} \quad i = {\mathcal{E}_k^{\text{tail}}} \\
		-1 \quad \text{if} \quad i = {\mathcal{E}_k^{\text{head}}} \\
		0 \quad \text{otherwise}
	\end{cases},
	\label{eq: B}
\end{equation}
where $\mathcal{E}_k^{\text{tail}}$ and $\mathcal{E}_k^{\text{head}}$ denote the tail and head nodes, respectively, of the edge $\mathcal{E}_k$, i.e., $\mathcal{E}_k = (\mathcal{E}_k^{\text{tail}},\mathcal{E}_k^{\text{head}})$.

A \emph{framework} is defined by the pair $(\mathbb{G}, p)$, where $p$ is the stacked vector of the agents' positions $p_i$ with $i\in\{1,\dots,n\}$. The stacked vector of the sensed relative positions by the agents can then be described by
\begin{equation}
	z = \overline B^Tp.
\end{equation}
Note that each vector $z_k = p_i - p_j$ in $z$ corresponds to the relative position associated with the edge $\mathcal{E}_k = (i, j)$.

\section{Distance-based daisy chain frameworks, mismatches and the uniform deployment on a line problem}
\label{sec: dai}
Assume that the agent's dynamics are described by the first-order model
\begin{equation}
	\dot p = u,
	\label{eq: pdyn1}
\end{equation}
where $u$ is the stacked vector of control inputs $u_i\in\R^2$ for $i=\{1,\dots,n\}$.

Consider the following incidence matrix defining a daisy-chain topology
\begin{equation}
B =
\begin{bmatrix}
	1 & 0 & \dots & 0 & 0 \\
	-1 & 1 & \dots & 0 & 0 \\
	\vdots & \vdots & \ddots & \vdots & \vdots \\
	0 & 0 & \dots & -1 &  1 \\
	0 & 0 & \dots & 0  &-1
\end{bmatrix},
\end{equation}
where $B\in\R^{|\mathcal{V}|\times(|\mathcal{V}|-1)}$. The incidence matrix that will help us to control the angles defined by the vectors $z_k$ and $z_{k+1}$ also corresponds to a daisy chain topology but $B_\theta\in\R^{(|\mathcal{V}|-1)\times(|\mathcal{V}|-2)}$, i.e., the first column of $B_\theta$ is related to $\theta_1$ as the angle between $z_1$ and $z_2$ and so on.

\subsection{Distance-based mismatched gradient-descent control}
For illustration, let us consider the case when our daisy chain framework consists of three agents. We then choose for the distance-based control the following potential function
\begin{equation}
	V(z) = \frac{1}{4}(||z_1||^2 - d_1^2)^2 + \frac{1}{4}(||z_2||^2 - d_2^2)^2,
	\label{eq: V}
\end{equation}
where $d_1$ and $d_2$ are the desired distances between the corresponding neighboring agents. Taking the gradient-descent of (\ref{eq: V}) (as used in \cite{Hector2016maneuvering}) we arrive at the following system
\begin{equation}
	\begin{cases}
	\dot p_1 &= -z_1e_1 \\
	\dot p_2 &=  z_1e_1 - z_2e_2 \\
	\dot p_3 &=  z_2e_2,
	\end{cases}
	\label{eq: grad}
\end{equation}
where $e_k = ||z_k||^2 - d_k^2, k\in\{1, 2\}$ are the distance error signals. Inspired by \cite{Hector2016maneuvering}, let us now include a distance mismatch $\mu_k \in\mathbb{R}$ in the edge $\mathcal{E}_k = (i,j)$, namely
\begin{equation}
d_k^{2\, \text{tail}} = d_k^{2\, \text{head}} - \mu_k,
	\label{eq: mu}
\end{equation}
where $d_k^{\text{tail}}$ and $d_k^{\text{head}}$ are the different desired distances that the agents $i$ and $j$ respectively in $\mathcal{E}_k=(i,j)$ want to maintain for the same edge. We consider that the mismatches are assigned to the second agent in (\ref{eq: grad}) such that we can arrive at the following expression
\begin{equation}
	\begin{cases}
	\dot p_1 &= -z_1e_1 \\
	\dot p_2 &=  z_1e_1 - z_2e_2 + \mu_1z_1 - \mu_2z_2 \\
	\dot p_3 &=  z_2e_2.
	\end{cases}
	\label{eq: prig}
\end{equation}
One can identify that system (\ref{eq: prig}) can be derived from a potential function as it has been done for system (\ref{eq: grad}) with the exception of the term $\mu_1z_1 - \mu_2z_2$. In fact, the gradient-descent-derived terms are responsible for the distance control between neighboring agents. If one drops all the terms in (\ref{eq: prig}) involving the control of the $d_k$'s, then one gets
\begin{equation}
	\begin{cases}
	\dot p_1 &= 0 \\
	\dot p_2 &=  \mu_1z_1 - \mu_2z_2 \\
	\dot p_3 &=  0.
	\end{cases}
	\label{eq: prig2}
\end{equation}
If one considers $\mu_1=\mu_2=c$ then system (\ref{eq: prig2}) follows precisely from the algorithm presented in \cite{kvinto2012equidistant} and in \cite{proskurnikov2016problem} for solving the problem of deployment on a line, i.e., two fixed points $p_1$ and $p_n$ defining a segment and the rest of agents will be deployed on such a segment at spots equally separated. In particular, as one will see in the following section, the algorithm is stable for $c\in\mathbb{R}^+$ and its compatibility with the distance-based gradient-descent control and its relation with biases in range sensors have been studied in \cite{ifacnew}.

\subsection{Deployment on a line problem}
In this section we will prove the stability of the algorithm introduced in system (\ref{eq: prig2}) for $\mu_1=\mu_2=c$ but for a general daisy chain topology. The stability analysis in this section is different from the one presented in \cite{kvinto2012equidistant} and in \cite{proskurnikov2016problem}. In particular, in this paper we analyze the derived error signals from the algorithm. This approach serves as a starting point for controlling polygons in the plane, not only collinear configurations. Let us define the following error vector
\begin{equation}
e_\theta = \overline B_\theta^T z,
	\label{eq: ez}
\end{equation}
where $e_\theta\in\R^{(|\mathcal{V}|-2)}$. Then the extension to $n$ agents from system (\ref{eq: prig2}) can be generalized as
\begin{equation}
	\begin{cases}
		\dot p_1 &= 0 \\
		\dot p_2 &= ce_{\theta_1} \\
		\vdots \\
		\dot p_{n-1} &= ce_{\theta_{n-2}} \\
		\dot p_n &= 0,
	\end{cases}
	\label{eq: prign}
\end{equation}
where $c\in\mathbb{R}^+$ is a constant gain. Let us write the dynamics of the signal $e_\theta(t)$. We first derive the dynamics of $z$ from system (\ref{eq: prign}), namely
\begin{equation}
	\dot z = \overline B^T\dot p = -c\overline B_\theta \overline B_\theta^T z= -c \overline B_\theta e_\theta,
\end{equation}
and noting that $\dot e_\theta = \overline B^T_{\theta}\dot z$ we have that
\begin{equation}
\dot e_\theta = -c \overline{B^T_\theta B_\theta}e_\theta.
	\label{eq: edyn}
\end{equation}

\begin{prop}
\label{pro: 1}
The origin of system (\ref{eq: edyn}) is globally exponentially stable. That is, all the agents from system (\ref{eq: prign}) will converge to a fixed point, namely $p(t) \to p^*$ as $t\to\infty$, where all the agents are equally spaced with respect to each other in a collinear fashion.
\end{prop}
\begin{pf}
Consider the following Lyapunov function $V = \frac{1}{2}||e_\theta||^2$, whose time derivative is given by
	\begin{equation}
		\frac{\mathrm{d}V}{\mathrm{dt}} = e^T_\theta \dot e^T_\theta = -c e_\theta^T \overline{B^T_\theta B_\theta}e_\theta.
	\end{equation}
	We know that $B_\theta$ defines a daisy chain topology, i.e., it does not contain any cycles, therefore the matrix $B^T_\theta B_\theta$ is positive definite (\cite{dimarogonas2008stability}). Hence the exponential stability of the origin of $e_\theta$ follows. Since the signal $e_\theta(t)$ converges exponentially fast to zero, we know $\overline B^T_\theta z(t) \to 0$ as $t\to\infty$, i.e., $z_k(t) - z_{k+1}(t) \to 0$ as $t\to\infty$. Thus, by observing system (\ref{eq: prign}), we have that $\dot p(t)$ also converges exponentially fast to zero. So one can conclude that $p(t)$ converges to a fixed point $p^*$ where all the agents are equally spaced and collinearly positioned.$\blacksquare$
\end{pf}
\begin{remark}
	Note that for the case $p_1(0) = p_n(0)$, all the agents will converge to the same point, i.e., $z(t)\to 0$ as $t\to\infty$.
\end{remark}
\subsection{Controlling relative magnitudes between relative positions}
The relative magnitude between two consecutive relative positions $z_k$ and $z_{k+1}$ can be trivially defined as $r_k z_k = r_{k+1}z_{k+1}$, where $r_k, r_{k+1}\in\mathbb{R}^+$ are the scaling factors that determine how the magnitude of one relative position with respect to its next neighboring one. This case encompasses, as in (\ref{eq: prig2}), the particular case of having all the agents equally spaced in the steady state, e.g., $r_k = 1, \forall k\{1, \dots, |\mathcal{E}|\}$. In particular, we have that
\begin{equation}
\tilde z = \overline D_r z,
	\label{eq: Dr}
\end{equation}
where $D_r \dfb \operatorname{diag}\left({\begin{bmatrix}r_1 & \dots & r_k \end{bmatrix}}\right),$ with $k\in\{1, \dots, |\mathcal{E}|\}$. So by redefining
\begin{equation}
	e_\theta = \overline B_\theta^T \tilde z, 
	\label{eq: rede}
\end{equation}
we have that the error dynamics derived from (\ref{eq: prign}), as we have done before in Proposition \ref{pro: 1}, are given by
\begin{equation}
\dot e_\theta = -c \overline{B^T_\theta D_r B_\theta}e_\theta,
\end{equation}
where the matrix $-B^T_\theta D_r B_\theta$ is Hurwitz since $D_r$ is a diagonal positive definite matrix. Therefore, the set defined by the vanishing of the signal (\ref{eq: rede}) is globally exponentially stable for system (\ref{eq: prign}).

\section{Controlling polygonal formations in the plane}
\label{sec: rot}
It is possible to extend the results of Proposition \ref{pro: 1} in order to deploy the team of agents on the plane in a more general way. We are going to show that by following the technique introduced in \cite{ifacnew} one is able to control the relative angle $\theta_k$ between two consecutive vectors $z_k$ and $z_{k+1}$. For formations where all these consecutive angles are equal to $\theta^*$, we provide a bound to such an angle in order to assert the (exponential) stability of the system. In particular, we will show that such a bound applies to the particular case of controlling regular polygons in the plane.

We introduce the consecutive angles $\theta_k$ to be controlled in the redefinition of the error signal $e_\theta$ as follows
\begin{equation}
	e_{\theta_k} = W\left(\frac{\theta_k}{2}\right)z_k - W\left(\frac{\theta_k}{2}\right)^Tz_{k+1}, \quad  \forall k\in\{1,\dots,|\mathcal{V}|-2\},
	\label{eq: etW}
\end{equation}
where $\theta_k \in (-\pi, \pi]$ and $W(\alpha)$ is the rotational matrix
\begin{equation}
	W(\alpha) = \begin{bmatrix}
	\cos(\alpha) & -\sin(\alpha) \\
	\sin(\alpha) & \cos(\alpha)
	\end{bmatrix}.
	\label{eq: W}
\end{equation}
Note that in (\ref{eq: etW}) we are comparing the clockwise rotated $z_k$ with the counterclockwise rotated $z_{k+1}$.

We now write in a compact form the stacked vector of errors in (\ref{eq: etW}) as
\begin{equation}
	e_\theta = B_{W}^T z,
	\label{eq: etheta}
\end{equation}
where
\small
\begin{equation}
	B_W = \begin{bmatrix}
		W\left(\frac{\theta_1}{2}\right) & 0 & \dots & 0 & 0 \\
		-W\left(\frac{\theta_1}{2}\right)^T & W\left(\frac{\theta_2}{2}\right) & \dots & 0 & 0 \\
	\vdots & \vdots & \ddots & \vdots & \vdots \\
		0 & 0 & \dots & -W\left(\frac{\theta_{n-1}}{2}\right)^T &  W\left(\frac{\theta_{n-2}}{2}\right) \\
		0 & 0 & \dots & 0  &-W\left(\frac{\theta_{n-2}}{2}\right)^T
\end{bmatrix}.
\end{equation}
\normalsize
Note that trivially $B_W$ is equal to $\overline B_\theta$, as defined in the end of Section \ref{sec: not}, if we set $\theta_k = 0, \forall k\in\{1,\dots,|\mathcal{V}|-2\}$. Therefore to deploy on a line is a particular case of the problem considered in this section.

Let us now write the dynamics of $z$ derived from system (\ref{eq: prign}) by employing the error signal (\ref{eq: etW})
\begin{equation}
\dot z = -c\overline B_\theta B_W^T z,
\end{equation}
so it allows us to derive the new error (linear) system dynamics given by
\begin{equation}
	\dot e_\theta = -c B_W^T \overline B_\theta e_\theta = -A(\theta)e_\theta,
	\label{eq: erW}
\end{equation}
where $\theta\in\mathbb{R}^{(|\mathcal{V}|-2)}$ is the stacked vector of all $\theta_k$ and $A(\theta)$ is shown in (\ref{eq: Atheta}) in the next page.
\begin{figure*}[!t]
\small
\begin{equation}
A(\theta) = \begin{bmatrix}
	W(\frac{\theta_1}{2})+W(\frac{\theta_1}{2})^T & -W(\frac{\theta_2}{2}) & 0 & \dots & 0 & 0 & 0 \\
	-W(\frac{\theta_2}{2})^T & W(\frac{\theta_2}{2})+W(\frac{\theta_2}{2})^T & -W(\frac{\theta_3}{2}) & \dots & 0 & 0 & 0 \\
	\vdots & \vdots & \vdots & \ddots & \vdots & \vdots & \vdots \\
	0 & 0 & 0 & \dots & -W(\frac{\theta_{n-3}}{2})^T & W(\frac{\theta_{n-3}}{2})+W(\frac{\theta_{n-3}}{2})^T & -W(\frac{\theta_{n-2}}{2}) \\
	0 & 0 & 0 & \dots & 0 & -W(\frac{\theta_{n-2}}{2})^T & W(\frac{\theta_{n-2}}{2})+W(\frac{\theta_{n-2}}{2})^T
\end{bmatrix}.\label{eq: Atheta}
\end{equation}
\hrulefill
\vspace*{4pt}
\normalsize
\end{figure*}
Note that now $A(\theta)$ is not positive definite in general. Therefore in order to check the stability of the origin of $e_\theta$ in (\ref{eq: erW}), one has to do an eigenvalue analysis for $A(\theta)$. 

Our numerical simulations have shown that not for all the values of $\theta$ the origin of (\ref{eq: erW}) is stable. In fact, the team of agents might converge to a different shape at the same time when they perform a steady-state motion. This effect has been shown not only for rigid formations with distance mismatches (see \cite{SMA16TACsub}), but also for \emph{flexible} formations (as in \cite{ifacnew}) as the daisy chain setup described in this paper. Nevertheless, we can provide an analytical result for the stability of a broad class of polygons where $\theta_k = \theta^*$. In particular we provide a bound for $\theta^*$ such that the formation is stable. Fortunately, this bound also applies to the set of regular polygons, which can be of interest in the field of formation control \cite{KrBrFr08}.
\begin{thm}
	\label{th: main}
	Consider the $n \geq 3$ agent system (\ref{eq: prign}) with $e_\theta$ defined as in (\ref{eq: etW}) and $\theta_k = \theta^* , \, \forall k\in\{1,\dots, n-2\}$ . Then, the origin of $e_\theta(t)$ in system (\ref{eq: erW}) is exponentially stable if and only if $|\theta^*| \leq \frac{2\pi}{n-1}$.
	\label{thm: t}
\end{thm}
\begin{pf}
Since all the 2D rotational matrices $W(\alpha)$ as in (\ref{eq: W}) commute, then $A(\theta)$ is unitarily similar to $\operatorname{diag}\{C(\theta), C(\theta)^\dagger\}$, where
\small
	\begin{equation}
		C(\theta) = \begin{bmatrix}
			2\cos\left(\frac{\theta^*}{2}\right) & -w(\theta^*) & &  \\
			-w(\theta^*)^\dagger & \ddots & \ddots & \\
			& \ddots & \ddots & -w(\theta^*) \\
			& & -w(\theta^*)^\dagger & 2\cos\left(\frac{\theta^*}{2}\right)
	\end{bmatrix},
	\end{equation}
\normalsize
	with $C(\theta)\in\R^{(n-2)\times (n-2)}, w(\theta) = e^{j\frac{\theta}{2}}$, $j$ is the imaginary unit, and the symbol $\dagger$ denotes for the complex conjugate transpose. The matrix $C$ is tridiagonal and Toeplitz, so its eigenvalues have the following analytical expression (\cite{noschese2013tridiagonal}):
	\begin{equation}
		\lambda_k(\theta^*) = 2\cos{\frac{\theta^*}{2}} + 2\cos\left(\frac{k\pi}{n-1}\right), \quad k\in\{1,\dots,n-2\}
	\end{equation}
	hence $C$ is positive definite (so the eigenvalues of $A(\theta)$ are positive) if and only if $|\theta^*| \leq \frac{2\pi}{n-1}$. Therefore the origin of $e_\theta(t)$ in system (\ref{eq: erW}) is exponentially stable. This fact implies the (exponential) convergence of system (\ref{eq: prign}) to the set given by $e_\theta = 0$ with $e_{\theta_k}$ as in (\ref{eq: etW}). $\blacksquare$
\end{pf}
\begin{remark}
	For the particular case of $n=3$, we have that for all the values of $\theta_1\in(-\pi,\pi]$ the system is stable, i.e., we are defining a triangle where its scale is determined by the constant positions $p_1(0)$ and $p_3(0)$.
\end{remark}
\begin{remark}
	Note that for regular polygons we have that $\theta^* = \pi - \frac{\pi(n-2)}{n}$ which satisfies the bound in Theorem \ref{thm: t}. The angle $\theta_k$ is not an inner angle of the polygon, but the angle between two consecutives $z_k$ and $z_{k+1}$.
\end{remark}
Note that the algorithm presented in (\ref{eq: prign}) with $e_\theta$ as in (\ref{eq: etheta}) is able to control 2D shapes employing ($n-2$) edges, which requires fewer than the ($2n - 3$) edges as in the gradient-descent control of the distance-based rigid setup. This is also the case if one employs position-based control (\cite{oh2015survey}). However, the algorithm presented in this paper has two important features that are not present in the position-based approach. First, the agents can work employing their own local frames of coordinates. Second, the steady-state orientation of the shape is not fixed, therefore allowing for rotational motion as we will show. These two advantages come from the fact that the presented algorithm is an extension of the mismatched distance-based formation control. 

\subsection{Controlling the scale of the prescribed shape}
Consider the example where six agents want to form an hexagon, so the four agents in the middle of the chain would control the \emph{inter-angles} $\theta_1, \dots, \theta_4 = \pi - \frac{2}{3}\pi$ and by looking at (\ref{eq: prign}) we notice that agents $1$ and $6$ are stationary. The idea is to apply the distance-based control to these two agents at the ends of the daisy chain, and therefore \emph{closing the chain}. For example, if we are controlling a regular polygon, then all the side-lengths are equal, i.e., $D_r$ is the identity matrix in (\ref{eq: Dr}). Therefore, if we control the distance $d$ between the first and the last agent, then the rest of distances between neighboring agents will also be equal to $d$.

For controlling the scale we assign the following control law, derived from a potential function like in (\ref{eq: V}), to agents $1$ and $n$
\begin{equation}
\begin{cases}
\dot p_1 &= -(p_n - p_1)(||p_n - p_1||^2 - d^2)  \\
\dot p_n &= (p_n - p_1)(||p_n - p_1||^2 - d^2)
\end{cases}.
\label{eq: prigdi}
\end{equation}
We have already noted that the convergence to the desired distance between agents $1$ and $n$ in the nonlinear system (\ref{eq: prigdi}) is exponential (\cite{sun2016exponential}). One can treat the terms (\ref{eq: prigdi}) as a disturbance that vanishes exponentially fast when they are into the dynamics of (\ref{eq: prign}), and hence the stability result in Theorem \ref{th: main} is not compromised and the scale of the formation can be controlled by only two agents. In fact, only one agent is needed if we set $\dot p_n = 0$ in (\ref{eq: prigdi}).

In case that one also desires to control the steady-state orientation of the formation, then a position-based control (with an exponential equilibrium) can be applied to agents $1$ and $n$.

\begin{remark}
	For the three agents example, the main difference of system (\ref{eq: prign}) with respect to the one presented in \cite{ifacnew} is that in the latter the sensing of $p_3 - p_1$ is not necessary for determining the scale of the triangular formation. In \cite{ifacnew}, the agents are also controlling the size of $z_k$ as in system (\ref{eq: prig}) where the gradient descent terms have not been dropped out.
\end{remark}

\subsection{Steering the prescribed shape in the plane}
\begin{figure}
\centering
\begin{tikzpicture}[line join=round]
\draw[dashed](1.5,0)--(0,0)--(0,1.5)--(1.5,1.5);
\draw[] (1.5,1.5)--(1.5,0);
\draw[](-.3,0)--(-.3,0.2)--(0,0.2);
\draw[](0.3,1.5)--(0.3,1.7)--(0,1.7);
\draw[dashed](0,0) -- (-1,0);
\draw[dashed](0,1.5) -- (0,2.5);
\draw[draw=red,arrows=->](0,0)--(.525,-.525);
\draw[draw=red,arrows=->](1.5,0)--(2.025,.525);
\draw[draw=red,arrows=->](1.5,1.5)--(.975,2.025);
\draw[draw=red,arrows=->](0,1.5)--(-.525,.975);
\draw[draw=blue,arrows=->](0,0)--(.75,0);
\draw[draw=blue,arrows=->](1.5,0)--(2.25,0);
\draw[draw=blue,arrows=->](1.5,1.5)--(2.25,1.5);
\draw[draw=blue,arrows=->](0,1.5)--(.75,1.5);
\filldraw(0,0) circle (2pt);
\filldraw(1.5,0) circle (2pt);
\filldraw(1.5,1.5) circle (2pt);
\filldraw(0,1.5) circle (2pt);
\draw[draw=black,arrows=->](.75,.75)--(1.125,.75);
\draw[draw=black,arrows=->](.75,.75)--(.75,1.125);
\draw[draw=black,arrows=->](-.75,1.0)--(-1.125,1.0);
\draw[draw=black,arrows=->](-.75,1.0)--(-.75,0.6);
	\node at (.95,.5) {\small $O_b$ \normalsize};\node at (-.75,1.2) {\small $O_g$ \normalsize};\node at (.75,-.2) {\small $z_1^*$ \normalsize};\node at (2.3,.75) {\small $||p_1-p_4||^*$ \normalsize}; \node at (0.75,1.7) {\small $z_3^*$ \normalsize};\node at (-.15,.75) {\small $z_2^*$ \normalsize};\node at (-.5,0.2) {\small $\theta_1^*$ \normalsize};\node at (.25,2) {\small $\theta_2^*$ \normalsize};
\end{tikzpicture}
	\caption{Explanation of control law (\ref{eq: mum}) for a square as a prescribed shape. The shape is achieved by the control of the relative angles $\theta_1$ and $\theta_2$ by agents $2$ and $3$ respectively to $\frac{\pi}{2} < \frac{2\pi}{3}$ rads, satisfying the bound in Theorem \ref{th: main}. Note that $\theta_1$ and $\theta_2$ do not define the inner angles of the polygon, but the angles between two consecutive $z_k$ and $z_{k+1}$. An inner angle is simply $\pi - \theta_k$. We set $D_r$ in (\ref{eq: Dr}) to the identity matrix, so all the norms $||z_k^*||$ will be equal at the steady-state. The scale of the square is determined by the control of $||p_1-p_4||$ (black solid). The velocity of an agent $\dot p_i^*$, at the desired shape, is the linear combination of the vectors from its associated relative positions. This velocity $\dot p_i^*$ can be decomposed into both translational (blue vectors) and rotational (red vectors) components. Note that these velocities are constant with respect to a frame of coordinates $O_b$ attached to the desired (body) shape.}
\label{fig: motion}
\end{figure}
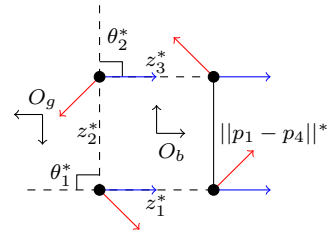

The exponential stability in Theorem \ref{th: main} can be further exploited. For example, one can employ the technique in \cite{Hector2016maneuvering} in order to steer the whole group with rotational and translational motions. It is obvious that a vector in the plane can be constructed as a linear combination of two non-parallel vectors. Therefore, agent $i$ can construct a velocity vector $\dot p_i^*$ by just combining two relative positions (available for the agent) from the formation. We note that this is indeed possible for agent $i$ from the system (\ref{eq: prign}) in combination with (\ref{eq: prigdi}). The main idea is to design a collection of steady-state velocities $\dot p_i^*$ by employing the relative positions such that $p \in\{z \, : \, (e_\theta = 0) \land (||p_1 - p_n|| = d)\}$ such that the desired shape is not destroyed, i.e., rigid body motions. For example, the control law introducing such an idea is given by
\begin{equation}
	\begin{cases}
		\dot p_1 &= -(p_n - p_1)(||p_n - p_1||^2 - d^2) \\
		& + \mu_{1_1}(p_n-p_1) + \mu_{1_2}z_1\\
		&\vdots \\
		\dot p_i &= ce_{\theta_{i+1}} + \mu_{i_1}z_{i-1} + \mu_{i_2}z_i \\
		&\vdots \\
		p_n &= (p_n - p_1)(||p_n - p_1||^2 - d^2) \\
		& + \mu_{n_1}p_n-p_1 + \mu_{n_2}z_{n-1} ,
	\end{cases}
	\label{eq: mum}
\end{equation}
where $i\in\{2, \dots, (n-1)\}$ and $\mu_{n_{\{1,2\}}}$ are the motion parameters responsible for the design of the velocities $\dot p_i^*$. We illustrate the physical meaning of (\ref{eq: mum}) in Figure \ref{fig: motion}.

 An algorithm describing how to compute these motion parameters such that they define rigid motions for generic shapes can be found in \cite{Hector2016maneuvering}. In fact, these motion parameters can be considered as a parametric disturbances for system (\ref{eq: prign}) considered in Theorem \ref{th: main}. In particular, it has been introduced in \cite{ifacnew}, inspired by the work in \cite{SMA16TACsub}, that the error-distance system defined by a rigid framework, whose equilibrium is the desired shape described by $\theta$, is autonomous and exponentially stable. Therefore, the stability of the error-distance system will not be compromised for small $\mu_{n_{\{1,2\}}}$'s (\cite{SMA16TACsub}), or for big control gains (\cite{Hector2016maneuvering}). This fact can be employed for giving bounds to the parameters $\mu_{n_{\{1,2\}}}$'s and the gain $c$ in (\ref{eq: mum}) in order to guarantee the exponential stability of the system for a set of desired velocities $\dot p_i^*$ (\cite{Hector2016maneuvering}).
\begin{remark}
	It is important to note that by the addition of the motion parameters in (\ref{eq: mum}) we might add undesired equilibria in the system. 
\end{remark}
\section{Simulations}
\label{sec: sim}
\begin{figure}
\centering
	\includegraphics[width=1\columnwidth]{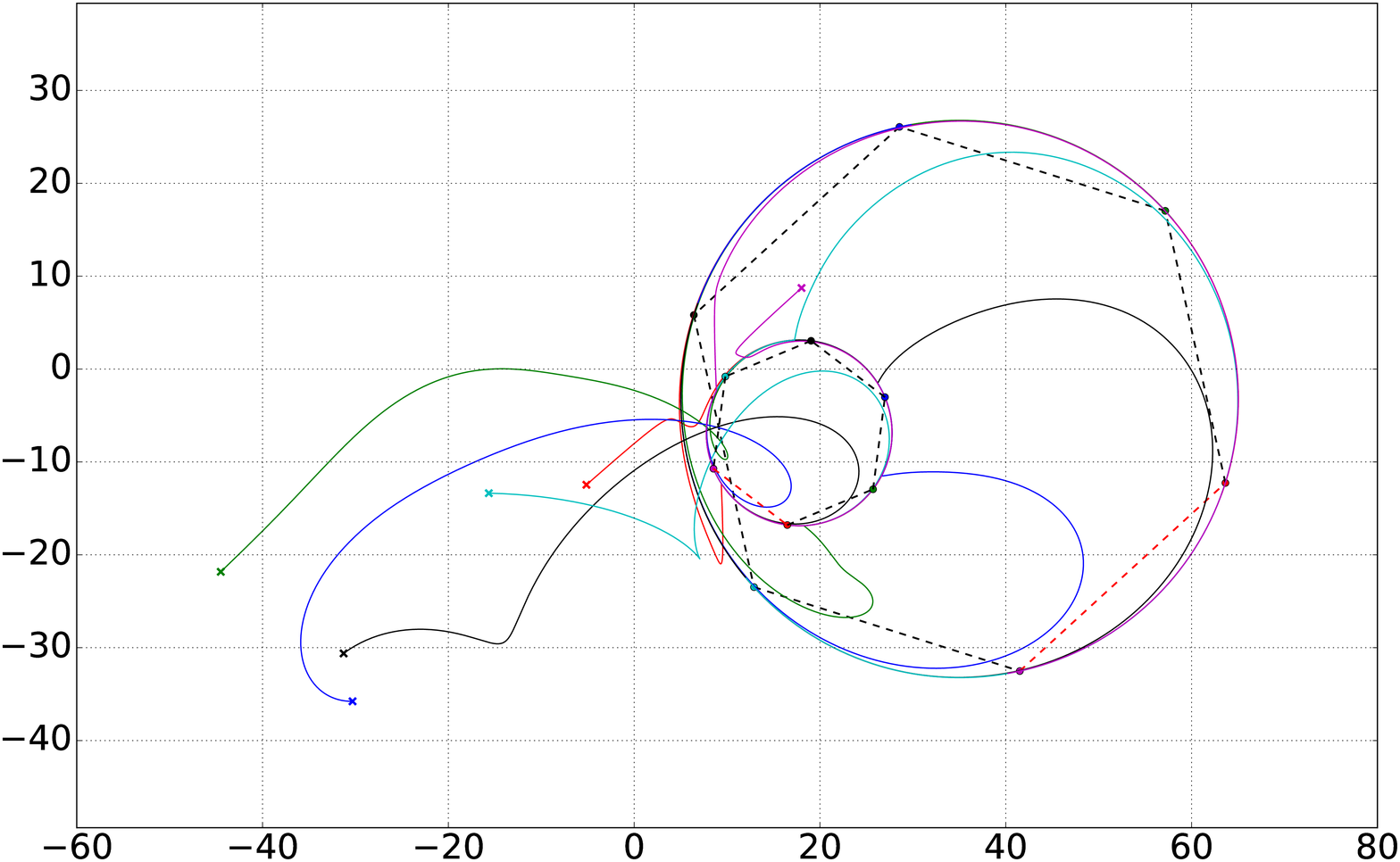} \\
	\includegraphics[width=0.45\columnwidth]{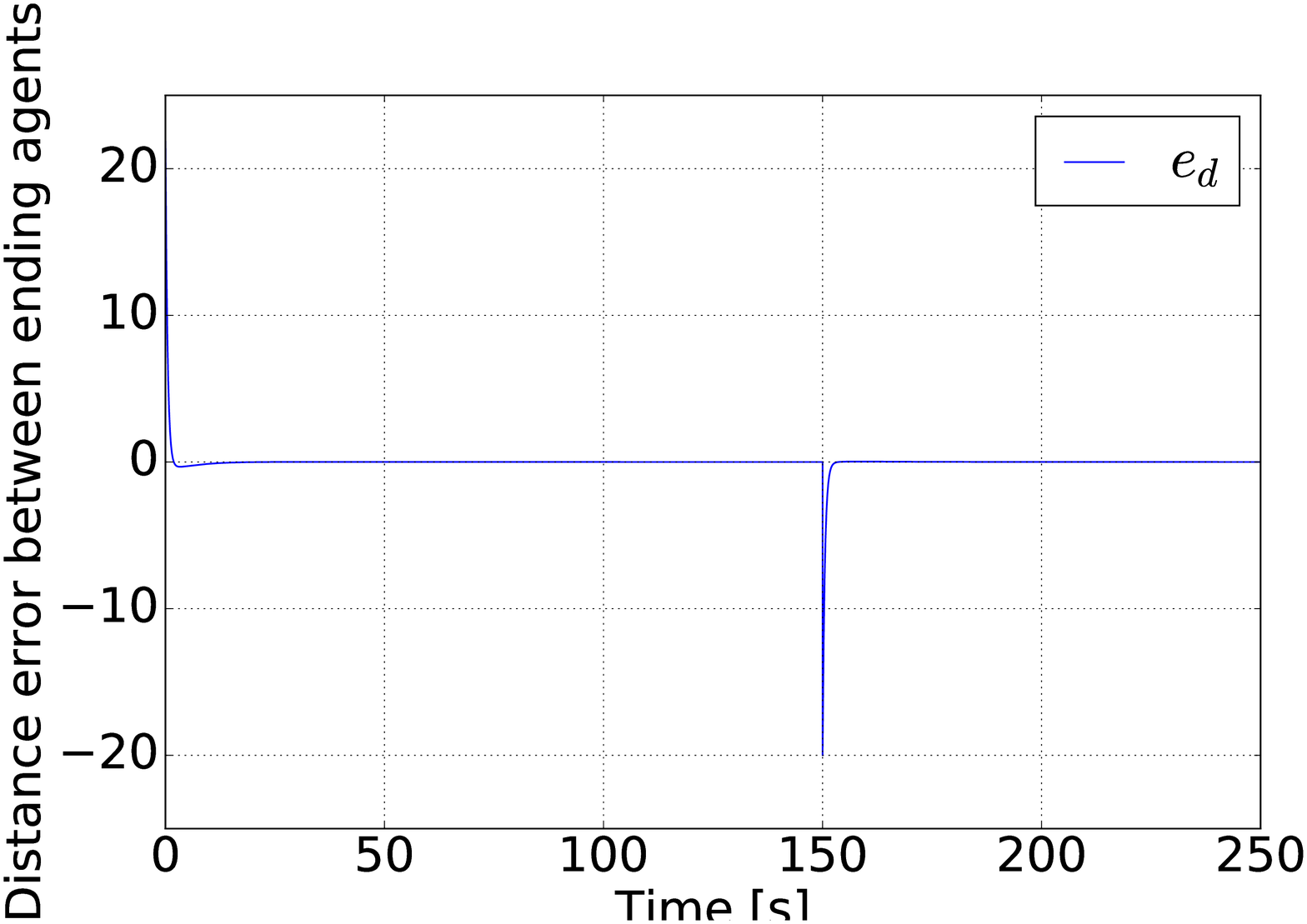}
	\includegraphics[width=0.45\columnwidth]{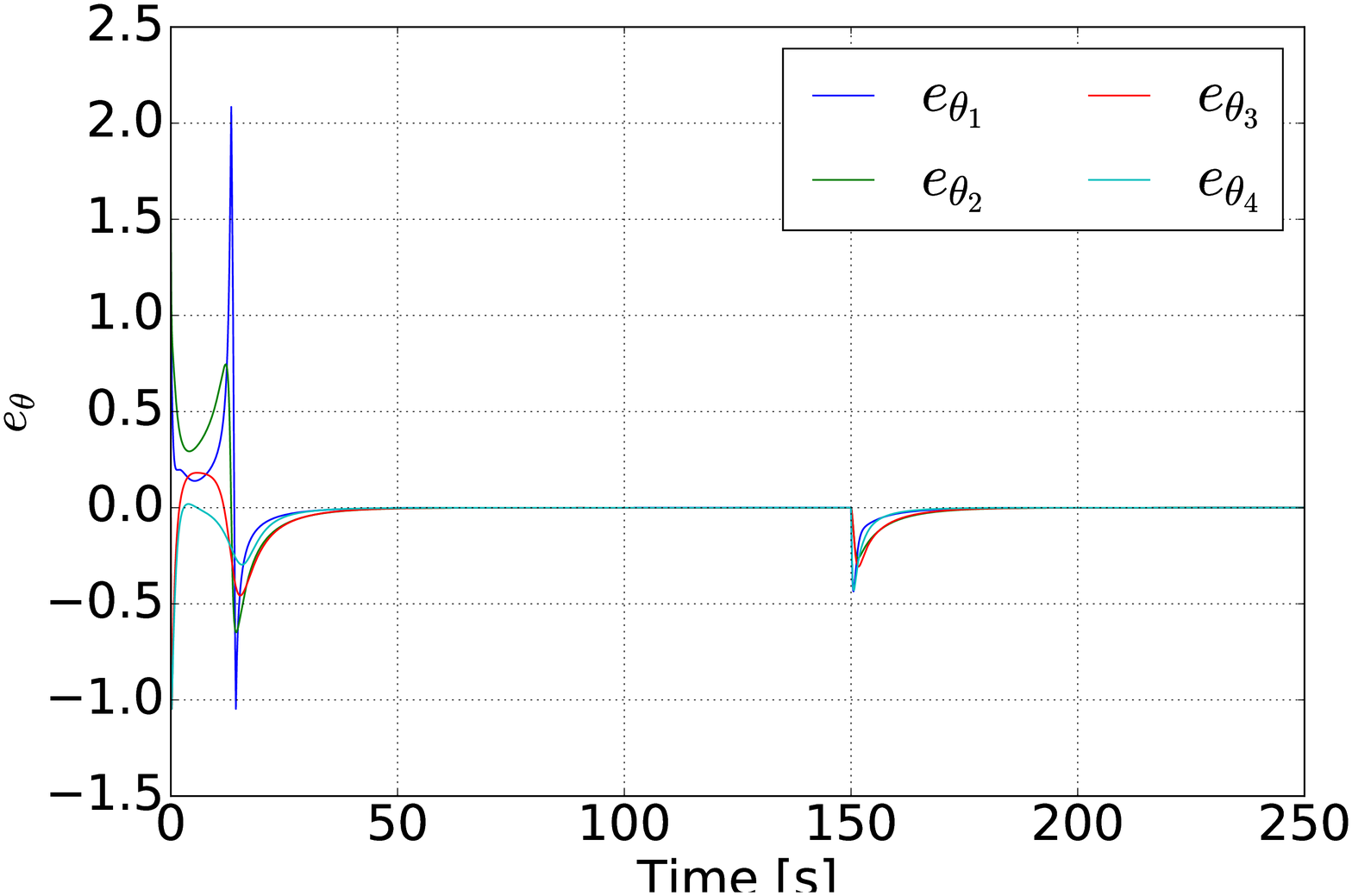}
	\caption{The figure on top shows the trajectories of the agents in solid
	color, where the crosses are the initial conditions. The red and magenta agents are the agents $1$ and $6$ respectively. The red dashed line corresponds to the controlled inter-distance between these two ending point agents in the daisy chain. Note how these two agents converge to the desired inter-distance $d$ (the side of the hexagon) describing almost a straight line before the rotational terms are dominant. The rest of the agents control the angles $\theta_k$ in order to achieve a regular hexagon. All the agents converge to a rotational motion about the centroid of the formed hexagon. This motion is given by setting all the motion parameters in (\ref{eq: mum}) to same constant. At time $t=150$ the agents $1$ and $6$ change the distance $d$ to be controlled to three times the starting one. This change results in a rescaling of the whole formation.}
\label{fig: hex}
\end{figure}

In this section we are going to validate the result of Theorem \ref{th: main} together with system (\ref{eq: mum}). We consider a team of six agents for achieving a regular hexagon, so $D_r$ is the identity matrix in (\ref{eq: Dr}). Since the inner angles of the hexagon are $\frac{2\pi}{3}$, we then set $\theta^* = \pi - \frac{2\pi}{3} = \frac{\pi}{3}$ which is smaller than the bound $\frac{2\pi}{5}$ given in Theorem \ref{th: main}. We define the error distance to be controlled by the agents $1$ and $6$ as $e_d \dfb ||p_1-p_6|| - d$ and we set $d=10$ in (\ref{eq: mum}). After time $t=150$ we set $d=30$. In addition we want to induce a spinning motion of the hexagon around its centroid. Such a motion can be accomplished by setting all the motion parameters in (\ref{eq: mum}) equal to $\mu = 0.025$. This can be checked by simple geometrical arguments or by employing the algorithm given in \cite{Hector2016maneuvering}. The simulation results are described in Figure \ref{fig: hex}.

\section{Conclusions}
\label{sec: con}
In this paper we have presented a distributed algorithm for controlling formations of a broad class of polygonal shapes, including regular ones, defined by a daisy chain topology. The first step is to construct an algorithm for deploying agents equally spaced on a line, which is derived from adding distance mismatches to a standard distance-based controller in the literature. Consequently, a number of properties, such as having agents working using their own local frames of coordinates and non-controlled steady-state orientation, are preserved. In the second step we show that with the addition of rotational matrices one can control the relative angle between two consecutive relative positions in the framework. It turns out that the desired shape is then exponentially stable for a broad class of polygons including regular ones. By exploiting this stability property, one can add a series of useful properties to the formation. Firstly, one can control the relative size of consecutive relative positions in the framework. Secondly, the scale of the whole shape can be achieved by only controlling the distance between the first and the last agent of the framework. Thirdly, motion parameters can be employed in order to steer the formation as a combination of translations and rotations. We are currently working to implement the proposed algorithm in a robotic testbed \cite{iros2017}.

\bibliography{nonrigid}             

\begin{thebibliography}{13}
\providecommand{\natexlab}[1]{#1}
\providecommand{\url}[1]{\texttt{#1}}
\providecommand{\urlprefix}{URL }
\expandafter\ifx\csname urlstyle\endcsname\relax
  \providecommand{\doi}[1]{doi:\discretionary{}{}{}#1}\else
  \providecommand{\doi}{doi:\discretionary{}{}{}\begingroup
  \urlstyle{rm}\Url}\fi

\bibitem[{Dimarogonas and Johansson(2008)}]{dimarogonas2008stability}
Dimarogonas, D.V. and Johansson, K.H. (2008).
\newblock On the stability of distance-based formation control.
\newblock In \emph{Decision and Control, 2008. CDC 2008. 47th IEEE Conference
  on}, 1200--1205. IEEE.

\bibitem[{Garcia~de Marina et~al.(2015)Garcia~de Marina, Cao, and
  Jayawardhana}]{MarCaoJa15}
Garcia~de Marina, H., Cao, M., and Jayawardhana, B. (2015).
\newblock Controlling rigid formations of mobile agents under inconsistent
  measurements.
\newblock \emph{Robotics, IEEE Transactions on}, 31(1), 31--39.

\bibitem[{Garcia~de Marina et~al.(2016{\natexlab{a}})Garcia~de Marina,
  Jayawardhana, and Cao}]{Hector2016maneuvering}
Garcia~de Marina, H., Jayawardhana, B., and Cao, M. (2016{\natexlab{a}}).
\newblock Distributed rotational and translational maneuvering of rigid
  formations and their applications.
\newblock \emph{Robotics, IEEE Transactions on}, 32(3), 684--697.

\bibitem[{Garcia~de Marina et~al.(2016{\natexlab{b}})Garcia~de Marina,
  Jayawardhana, and Cao}]{HectorCDC16}
Garcia~de Marina, H., Jayawardhana, B., and Cao, M. (2016{\natexlab{b}}).
\newblock Distributed scaling control of rigid formations.
\newblock In \emph{Proceedings of the Decision and Control Conference, 2016.
  CDC 2016. 55th IEEE Conference on}. IEEE.

\bibitem[{Garcia~de Marina et~al.(2017)Garcia~de Marina, Siemonsma,
  Jayawardhana, and Cao}]{iros2017}
Garcia~de Marina, H., Siemonsma, J., Jayawardhana, B., and Cao, M. (2017).
\newblock Design and implementation of formation control algorithms for fully
  distributed multi-robot systems.
\newblock In \emph{Intelligent Robots and Systems (IROS), 2017 IEEE/RSJ
  International Conference on, Submitted}.

\bibitem[{Garcia~de Marina and Sun(2017)}]{ifacnew}
Garcia~de Marina, H. and Sun, Z. (2017).
\newblock Controlling a triangular flexible formation of autonomous agents.
\newblock In \emph{Proceedings of the 2017 IFAC World Congress}. IFAC.

\bibitem[{Krick et~al.(2009)Krick, Broucke, and Francis}]{KrBrFr08}
Krick, L., Broucke, M.E., and Francis, B.A. (2009).
\newblock Stabilization of infinitesimally rigid formations of multi-robot
  networks.
\newblock \emph{International Journal of Control}, 82, 423--439.

\bibitem[{Kvinto and Parsegov(2012)}]{kvinto2012equidistant}
Kvinto, Y.I. and Parsegov, S. (2012).
\newblock Equidistant arrangement of agents on line: Analysis of the algorithm
  and its generalization.
\newblock \emph{Automation and Remote Control}, 73(11), 1784--1793.

\bibitem[{Mou et~al.(2016)Mou, Belabbas, Morse, Sun, and
  Anderson}]{SMA16TACsub}
Mou, S., Belabbas, M.A., Morse, A.S., Sun, Z., and Anderson, B.D.O. (2016).
\newblock Undirected rigid formations are problematic.
\newblock \emph{IEEE Transactions on Automatic Control}, 61(10), 2821--2836.

\bibitem[{Noschese et~al.(2013)Noschese, Pasquini, and
  Reichel}]{noschese2013tridiagonal}
Noschese, S., Pasquini, L., and Reichel, L. (2013).
\newblock Tridiagonal toeplitz matrices: properties and novel applications.
\newblock \emph{Numerical Linear Algebra with Applications}, 20(2), 302--326.

\bibitem[{Oh et~al.(2015)Oh, Park, and Ahn}]{oh2015survey}
Oh, K.K., Park, M.C., and Ahn, H.S. (2015).
\newblock A survey of multi-agent formation control.
\newblock \emph{Automatica}, 53, 424--440.

\bibitem[{Proskurnikov and Parsegov(2016)}]{proskurnikov2016problem}
Proskurnikov, A.V. and Parsegov, S. (2016).
\newblock Problem of uniform deployment on a line segment for second-order
  agents.
\newblock \emph{Automation and Remote Control}, 77(7), 1248--1258.

\bibitem[{Sun et~al.(2016)Sun, Mou, Anderson, and Cao}]{sun2016exponential}
Sun, Z., Mou, S., Anderson, B.D.O., and Cao, M. (2016).
\newblock Exponential stability for formation control systems with generalized
  controllers: A unified approach.
\newblock \emph{Systems \& Control Letters}, 93, 50--57.

\end{thebibliography}

\end{document}